\documentclass[prb,twocolumn,amsmath,amssymb,superscriptaddress]{revtex4-1}
\usepackage[utf8x]{inputenc}
\usepackage{graphicx}
\usepackage{psfrag}
\usepackage{dcolumn}
\usepackage{bm}
\usepackage{color}

\begin{document}

\preprint{APS/123-QED}

\title{Quantum phase transitions from analysis of the polarization amplitude}

\author{Bal\'azs Het\'enyi}
\affiliation{Department of Theoretical Physics and MTA-BME Exotic
  Quantum Phases ``Momentum'' Research Group, Budapest University of Technology and Economics, 1521  Budapest, Hungary}
\affiliation{Department of Physics, Bilkent University,  TR-06800 Bilkent, Ankara, Turkey}
\author{Bal\'azs D\'ora}
\affiliation{Department of Theoretical Physics and MTA-BME Lend\"ulet Topology and Correlation Research Group,
Budapest University of Technology and Economics, 1521 Budapest, Hungary}

\begin{abstract}
In the modern theory of polarization, polarization itself is given by
a geometric phase.  In calculations for interacting systems the
polarization and its variance are obtained from the polarization
amplitude.  We interpret this quantity as a discretized characteristic
function and derive formulas for its cumulants and moments.  In the
case of a non-interacting system, our scheme leads to the
gauge-invariant cumulants known from polarization theory.  We study
the behavior of such cumulants for several interacting models.  In a
one-dimensional system of spinless fermions with nearest neighbor
interaction the transition at which gap closure occurs can be clearly
identified from the finite size scaling exponent of the variance.
When next nearest neighbor interactions are turned on a model with a
richer phase diagram emerges, but the finite size scaling exponent is still
an effective way to identify the localization transition.
\end{abstract}

\maketitle

\section{Introduction}

In crystalline systems the polarization is
expressed~\cite{King-Smith93,Resta94} as a Berry
phase,~\cite{Berry84,Xiao10} more specifically, its variant which
arises when crossing the Brillouin zone, the Zak phase,~\cite{Zak89}
rather than in terms of an ordinary observable.  This quantity is also
the starting point in deriving topological
invariants.~\cite{Bernevig13,Thouless82,Thouless83,Fu06} The Zak phase
itself corresponds to the first member in a series of gauge-invariant
cumulants (GIC), first studied by Souza, Wilkens, and
Martin~\cite{Souza00} (SWM).  While in band structure calculations one
can simply discretize~\cite{King-Smith93} the integrals over the
Brillouin zone, in interacting systems the polarization is obtained
from the expectation value of the momentum shift
operator,~\cite{Resta98,Resta99} also known as the polarization
amplitude.  From this quantity Resta and
Sorella~\cite{Resta98,Resta99} derived the polarization itself and its
variance.  The latter has been used
extensively~\cite{Aligia99,Nakamura02,Oshikawa03} as a localization
criterion~\cite{Kohn64} for the metal insulator transition.  For
higher order cumulants, expressions in the spirit of
Refs. \onlinecite{Resta98} and \onlinecite{Resta99} have not been
derived.  For ordinary expectation values higher order moments and
cumulants enable finite size
scaling.~\cite{Fisher72,Binder81,Binder92}

Some
studies,~\cite{Aligia99,Nakamura02,Oshikawa03,Kobayashi18,Oshikawa18,Nakamura18}
focus on the properties of the total momentum and total position shift
operators.  Nakamura and Voit~\cite{Nakamura02} showed the relation
between the Lieb, Schultz, and Mattis argument~\cite{Lieb61} and the
work of Resta and Sorella.~\cite{Resta98,Resta99} as well as
calculated renormalization group flows based on sine-Gordon theory.
Oshikawa found~\cite{Oshikawa18} a topological relation between
commensurability and conductivity using the total momentum and total
position shift operators, relating the number of low-lying states of
an insulator when the filling is an irreducible fraction $p/q$.  It is
also possible to derive~\cite{Hetenyi13} a topological invariant for
the Drude weight using the shift operators.  Closed expressions for
the finite size scaling exponent of $F_q$ were derived and calculated
numerically for a set of canonical models by Kobayashi {\it et
  al.}~\cite{Kobayashi18} It was also shown that definite scaling
relations apply to $F_q$ in some regions of the {\it metallic} state
of a strongly correlated model.  Recently there has also been an
interest~\cite{Patankar18,Yahyavi17} in the study of higher order
cumulants of the polarization.  Patankar {\it et
  al.}~\cite{Patankar18} showed that the third cumulant of the
polarization corresponds to the shift current, which gives the
nonlinear response in second harmonic generation experiments, which
show that this quantity exhibits a characteristic enhancement in Weyl
semimetals.~\cite{Patankar18}

In this paper, we give the discrete formulas for the cumulants and
moments based on the polarization amplitude up to any order.  We
construct the quantities relevant to finite size scaling, and show
that it is possible to locate phase transition points the usual way.
$F_q$ is interpreted as a characteristic function, and discrete
derivative approximations with respect to $q$ are applied to obtain
expressions for the gauge-invariant cumulants.  In contrast to
Kobayashi {\it et al.}~\cite{Kobayashi18} we focus on the moments
derived from $F_q$ rather than $F_q$ itself.  The moments and
cumulants seem to us physically more tangible as physical quantities,
more importantly their scaling turns out to be sensitive to the
metal-insulator transition, even though the leading scaling exponent
found in Ref. ~\onlinecite{Kobayashi18} cancels in our construction.
We also sketch the proof that when our construction is applied to a
non-interacting system the cumulants correspond to the GICs studied in
SWM.~\cite{Souza00} We also derive the connection between the GICs and
the probability distribution arising from the Wannier function of a
given system.

We calculate the cumulants for models of spinless interacting fermions
which exhibit a variety of phase transitions.  When only nearest
neighbor interactions are present a Luttinger liquid (LL) to
charge-density wave (CDW) transition occurs in the regime of positive
interaction, and a metallic state to phase separation when $V$ is
negative.  Even for small system sizes, the transition points are very
well located.  The variance scales as the square of the system size in
the metallic phase, expected based on comparing the variance of the
polarization with the polarizability.~\cite{Chiappe18,Baeriswyl00} We
also construct the analog of the Binder
cumulant,~\cite{Binder81,Binder92} a quantity which is particularly
sensitive to size effects around phase transitions.  In the metallic
regions, these quantities show critical behavior.

\begin{figure}[ht]
 \centering
 \includegraphics[width=\linewidth,keepaspectratio=true]{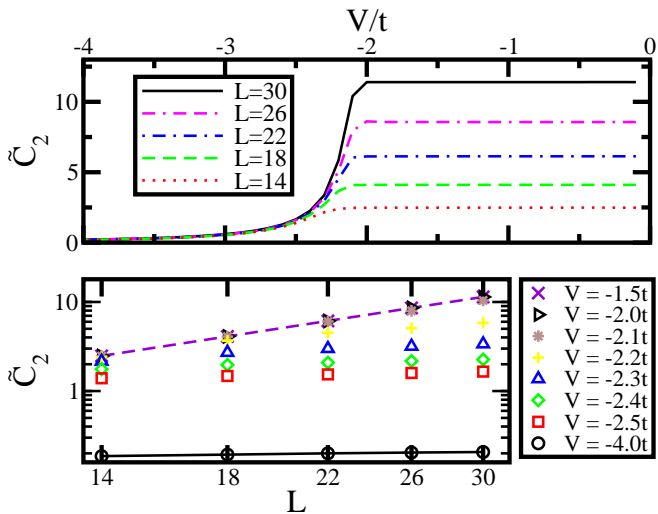}
 \caption{(Color online) Upper panel: $\tilde{C}_2$ as a function of
   $V/t$ for different system sizes.  Lower panel: $\tilde{C}_2$ as a
   function of system size for different values of $V$ on a log-log
   plot.  The two lines indicate fits of the function $f(L) = a
   L^\gamma + b$ for the two cases $V=-4.0t$ and $V=-1.5t$.}
 \label{fig:VmM2}
\end{figure}

\section{Discrete formulas for moments and cumulants}

For a system periodic in $L$ we first define the quantity
\begin{equation}
  \label{eqn:Fq}
  F_q = \langle \Psi |\exp(i 2 \pi q \hat{X} /L)| \Psi \rangle,
\end{equation}
where $\hat{X} = \sum_{j=1}^N x_j \hat{n}_j$ .  In terms of $F_q$
the $n$th moment can be written as
\begin{equation}
  \label{eqn:Mn}
  M_n = \left(\frac{L}{2 \pi i}\right)^n
 \left[ F_q
 \right]_{q=0}^{(n)}
\end{equation}
or the $n$th cumulant as
\begin{equation}
  \label{eqn:Cn}
  C_n = \left(\frac{L}{2 \pi i}\right)^n
 \left[ \ln F_q
 \right]_{q=0}^{(n)}.
\end{equation}
In the above equations the notation $[f_q]_{q=0}^{(n)}$ means discrete
derivative (finite difference) of order $n$ of the function $f_q$ at
$q=0$.  In a periodic system $q$ take only integer values.  Note that
Eqs. (\ref{eqn:Mn}) and (\ref{eqn:Cn}) amount to interpreting the
quantity $F_q$ as a discretized characteristic function.  It is easily
verified that the Resta~\cite{Resta98} and
Resta-Sorella~\cite{Resta99} formulas for the first and second
cumulants, respectively, are reproduced from Eq. (\ref{eqn:Cn}).  Zak
also wrote~\cite{Zak00} an expression for the polarization, which
corresponds to $M_1$ in a symmetric finite difference approximation.

Below we calculate cumulants by first obtaining the total position
$C_1$ and redefining $F_q$ as follows
\begin{equation}
  F_q = \langle \Psi |\exp(i 2 \pi q \hat{X} /L)| \Psi \rangle \exp(-i
  2 \pi q C_1 /L).
\end{equation}
This step is a mere a shift in the coordinate system, and is for
numerical convenience.  (Cumulants of order greater than one are
independent of the average.)  We then take the derivative of $\ln F_q$
with respect to $F_q$ analytically, resulting in a sum of products of
moments, and then express the moments via discrete derivatives.  For
example, the second cumulant is
\begin{equation}
 \tilde{C}_2 = M_2,
\end{equation}
where $M_2$ is are given by Eq. (\ref{eqn:Mn}).  $M_1$ is zero due to
the shift by $C_1$.  The finite difference derivative expressions in
this case are correct up to $\mathcal{O}(L^{-2})$.

\begin{figure}[ht]
 \centering
 \includegraphics[width=\linewidth,keepaspectratio=true]{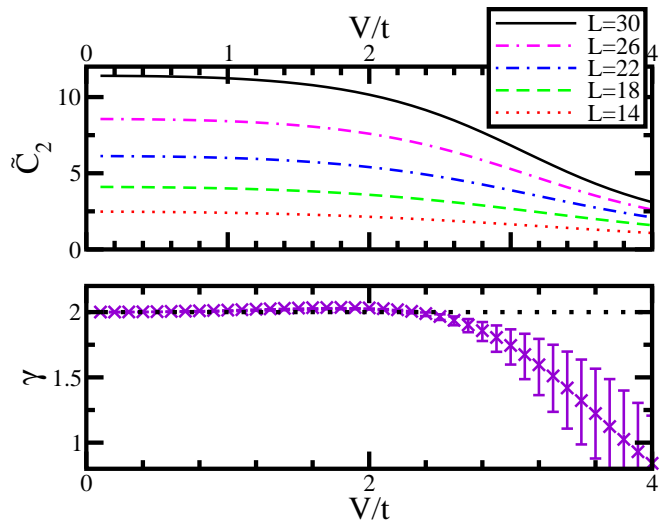}
 \caption{(Color online) Upper panel: $\tilde{C}_2$ as a function of
   $V/t$ for different system sizes.  Lower panel: scaling exponent
   $\gamma$ as a function of $V/t$.  The black dotted line indicates
   the value of two.  The error bars were multiplied by ten to make
   them more visible.}
 \label{fig:VM2}
\end{figure}

We now show that in a non-interacting framework
Eqs.(\ref{eqn:Fq}) to (\ref{eqn:Cn}) reproduce the GICs derived by
SWM.~\cite{Souza00}  We also derive the criterion which connects
Eq. (\ref{eqn:Fq}) with the characteristic function of the
distribution corresponding to the Wannier function, also in a
non-interacting system.  The derivation is based on the work of
Resta.~\cite{Resta98}

Consider a crystalline system of lattice constant $a$ with periodic
boundary conditions over $M$ cells ($L=Ma$), leading to $M$ equally
spaced Bloch vectors,
\begin{equation}
  k_s = \frac{2\pi}{M a} s,\hspace{1cm}s=0,...,M-1.
\end{equation}
The Bloch functions take the form
\begin{equation}
  \psi_{k_s,m}(x) = \exp(i k_s x) u_{k_s,m}(x),
\end{equation}
where $u_{q_x,m}(x)$ is Bloch function periodic in $a$, and $m$ is a
band index.  There are $N/M$ occupied bands in the ground state wave
function, which can be written
\begin{equation}
  \Psi_0 = {\bf A} \prod_{m=1}^{N/M} \prod_{s=0}^{M-1}\psi_{k_s,m},
\end{equation}
where ${\bf A}$ is the antisymmetrizer.  We now evaluate $F_q$ for
this wave function,
\begin{equation}
  F_q = \langle \Psi_0 |\exp(i 2 \pi q \hat{X} /L)| \Psi_0 \rangle =
  \mbox{det} S,
\end{equation}
where 
\begin{equation}
  S_{s m,s' m'} = \int_0^L dx \hspace{.1cm} \psi^*_{k_s,m}(x) \exp \left(i \frac{2 \pi }{L} q x \right) \psi_{k_{s'},m'}(x).
\end{equation}
We used the fact that the overlap of determinants equals the
determinant of overlaps.  Due to the orthogonality properties of the
Bloch wave functions $S_{s m,s' m'}$ is only finite if $s = s' + q$,
and $F_q$ determinant becomes
\begin{equation}
  \label{eqn:Fqband}
  F_q = \prod_{s=0}^{M-1}  \mbox{det} S(k_s,k_{s+q}).
\end{equation}
where
\begin{equation}
  S_{m,m'}(k_s,k_{s+q}) = \int_0^L dx \hspace{.1cm} \psi^*_{k_s,m}(x) \exp \left(i \frac{2 \pi }{L} q x \right) \psi_{k_{s+q},m'}(x).
\end{equation}
In terms of the periodic Bloch functions, the matrix $S(k_s,k_{s+q})$
becomes
\begin{equation}
  S_{m,m'}(k_s,k_{s+q}) = \int_0^L dx \hspace{.1cm} u^*_{k_s,m}(x) u_{k_{s+q},m'}(x).
\end{equation}
The quantity $F_q$ is the moment generating function, whereas $\ln
F_q$ is the cumulant generating function.  Taylor expanding $\ln F_q$
in $k_{s+q}-k_s = 2\pi q/L$ and taking the limit $L\rightarrow \infty$
gives the GICs of SWM (see Eqs. (32) and (33) of
Ref. \onlinecite{Souza00}).  For example, taking the second derivative of
$F_q$ in Eq. (\ref{eqn:Fqband}) and the limit $L\rightarrow \infty$ we
obtain
\begin{equation}
  C_2 = - \frac{1}{2\pi}\int_{BZ} dk (\langle u_{k,m} |\partial^2_k| u_{k,m}\rangle - \langle u_{k,m} |\partial_k| u_{k,m}\rangle^2).
\end{equation}

We also derive the condition under which the moments or GICs
correspond to the true moments or cumulants of the total position in a
band system.  The same derivation was used by Zak~\cite{Zak89} to show
that the Zak phase corresponds to the expectation value of the
position over the Wannier function of a given band.  Using the
definition of the Wannier function
\begin{equation}
  u_{k,m}(x) = \sum_{p=-\infty}^\infty \exp\left(ik(p a - x)\right) a_m (x - p a),
\end{equation}
we can rewrite the matrix elements as
\begin{eqnarray}
  S_{m,m'}(k_s,k_{s+q}) =  \sum_{p=-\infty}^\infty  \sum_{p'=-\infty}^\infty \int_0^L dx \\
  \exp\left(-i k_s (p a - x)\right) a^*_m (x - p a) \nonumber \\
  \exp\left(i k_{s+q}(p' a - x)\right) a_{m'} (x - p' a). \nonumber
\end{eqnarray}
We can extend the range of the integral to infinity and after further
rearrangements obtain
\begin{eqnarray}
  S_{m,m'}(k_s,k_{s+q}) =  \sum_{\Delta p=-\infty}^\infty  \exp\left(-i k_s \Delta p a \right)  \\
\int_{-\infty}^\infty dx \hspace{.1cm} a^*_m(x - \Delta p a)a_{m'}(x)
  \exp\left(- i \frac{2\pi}{Ma}q x\right) , \nonumber
\end{eqnarray}
where $\Delta p = p - p'$.  Assuming that the overlap between Wannier
functions centered in different unit cells is negligible, the matrix
elements become
\begin{equation}
  S_{m,m'}(k_s,k_{s+q}) = \int_{-\infty}^\infty dx \hspace{.1cm}
  a^*_m(x)a_{m'}(x) \exp\left(- i \frac{2\pi}{Ma}q x\right).
\end{equation}
In this case $F_q = \prod_{s=0}^{M-1} \mbox{det} S(k_s,k_{s+q})$ is
the characteristic function of the squared modulus of the determinant
of Wannier functions corresponding to occupied bands.  Note that a
similar approximation was derived in Ref. \onlinecite{Yahyavi17}.

The moments derived above were used to construct the maximally
localized Wannier functions~\cite{Marzari97,Marzari12} by optimizing
the variance of the position.  For a non-interacting system, when the
thermodynamic limit is taken, the resulting variance (constructed out
of moments) is {\it not} gauge invariant.  However, one can apply an
arbitrary phase to the full {\it many-body} wave function in
Eqs. (\ref{eqn:Fq}) and (\ref{eqn:Mn}) without changing $M_n$.  In
other words, the lack of gauge invariance manifests in the case of
separable wave functions, for example, product states of
single-particle wave functions, when individual orbitals can take
arbitrary phases.

\section{Interacting model of spinless fermions with nearest neighbor interaction}

We study an interacting model of spinless fermions on a lattice in one
dimension with Hamiltonian
\begin{equation}
  H = \sum_{i=1}^L [-t(c^\dagger_{i+1} c_i + c^\dagger_i c_{i+1}) + V n_i n_{i+1} ].
\end{equation}
We solve this system via exact diagonalization at half-filling.  Our
calculations include systems with periodic boundary conditions at
half-filling with an odd number of particles (the ground state
according to the Perron-Frobenius theorem).  At half filling, this
model exhibits a transition at $V=2t$ and at $V=-2t$.  The former is a
continuous transition between a LL at small $V/t$ and a CDW at large
$V$, the latter is first-order.  For $2t>-V$ the system is also LL,
while for $2t<-V$ the ground state is phase separated with particles
tending to cluster near each other.  In the large $V$ limit the ground
state is one in which all particles form a single cluster.  This state
is highly degenerate, such a state can be displaced by an arbitrary
number of sites resulting in states with the same energy.

\subsection{Finite size scaling of cumulants}

$\tilde{C}_2$ is shown in Fig. \ref{fig:VmM2} for the attractive case.
The upper panel shows this quantity as a function of $V/t$, while the
lower one for fixed values of $V/t$ as a function of system size.  In
the upper panel, the cumulant increases until the phase transition
point, but then levels off to a constant value in the conducting
phase.  The scaling exponent is difficult to determine in a region
close $V=-2t$ in the insulating phase, but it is $\gamma=0.0$ far from
the transition point (see line fit to data points for $V=-4.0t$ in
Fig. \ref{fig:VmM2}), and it gives a value of $\gamma=2.0$ (see line
fit to data points for $V=-1.5t$ in Fig. \ref{fig:VmM2}) in the
conducting phase.  Even for values of $V$ closer to the transition
point, $\tilde{C}_2$ tends to level off to a constant value for larger
system sizes, indicating a scaling exponent of $\gamma=0.0$ (see
Fig. \ref{fig:VmM2}, data for $V=-2.5t$, $V=-2.4t$).

\begin{figure}[ht]
 \centering
 \includegraphics[width=\linewidth,keepaspectratio=true]{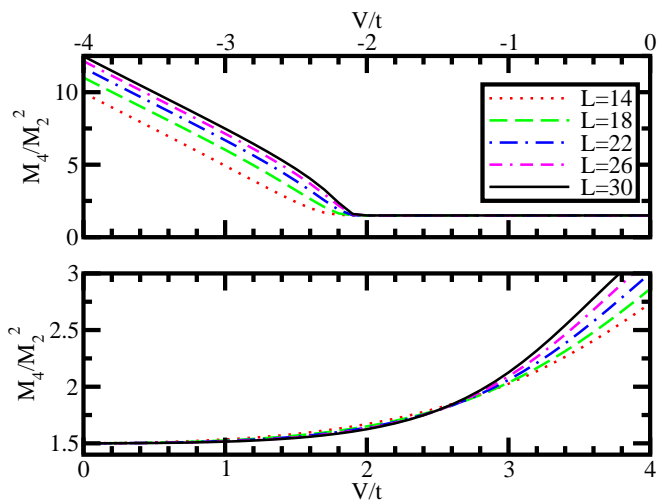}
 \caption{Upper panel(Lower panel): moment ratio $M_4/M_2^2$ for the
   attractive(repulsive) case.}
 \label{fig:U4}
\end{figure}

On the repulsive side, the behavior of these quantities is depicted in
Fig. \ref{fig:VM2}.  In the upper panel, $\tilde{C}_2$ is shown.  As
$V$ increases, the cumulant decreases monotonically, even in the
conducting phase ($V/t<2$).  What is remarkable is that, inspite of
this, the scaling exponent (lower panel) is very close to a value of
$2$ throughout the conducting phase, and starts to decrease as a
function of $V$ in the insulating phase.  The error bars for the
scaling exponent are $\mathcal{O}({10^{-4}})$ in the LL phase,
increase around the known phase transition point by several orders of
magnitude, until $V/t=3.7$ where they reach a maximum, and start to
decrease, indicative of the significant shifting and smearing of the
KT transition for these system sizes.

We can connect the fact that in the metallic phase
the scaling exponent of the second cumulant with $\gamma$ is two, and
decreases when the system enters the insulating phase.  It is
well-known~\cite{Chiappe18} that the polarizability obeys precisely
this scaling behavior, and the second cumulant gives an upper
bound~\cite{Baeriswyl00} to the polarizability.

\begin{figure}[ht]
  \centering
  \psfrag{x}{$x$}
  \psfrag{F}{$F$}
  \psfrag{V/t}{$V/t$}
   \includegraphics[width=\linewidth,keepaspectratio=true]{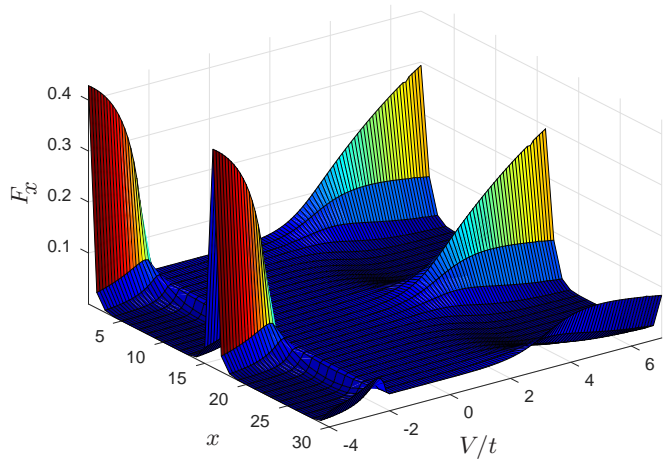}
 \caption{(Color online) Fourier transform of $F_q$ as a function of
   $x$ and the coupling constant $V/t$.}
 \label{fig:Px}
\end{figure}

Comparing with the results of Ref. \onlinecite{Kobayashi18} we see the
advantage of using the cumulants for thermodynamic scaling.  There it
was determined that $F_q \approx L^{\beta(V)}$ for this model, meaning
that the scaling of $F_q$ depends on the interaction.  However, the
scaling of $\tilde{C}_2$ within the metallic phase is independent of
$V$.  The scaling exponent $\beta(V)$ is the leading scaling exponent
in $F_q$ which was found to be linear in $q$ for a number of models.
In our definition of the cumulant, we take the derivative of $\ln F_q$
analytically with respect to $F_q$ (which still results in derivatives
of $F_q$).  For $\tilde{C}_2$ two derivatives in $q$ make $\beta{V}$
disappear, so the scaling we find is unaffected by it.  We find the
expected scaling of the variance of the total position in the LL phase
(Figs. \ref{fig:VmM2} and \ref{fig:VM2}) throughout the entire
metallic region, while in the attractive region
Ref. \onlinecite{Kobayashi18} reports definite scaling only for $-t < V <0$.

One way to apply the finite size scaling hypothesis~\cite{Fisher72} to
critical phenomena makes use of the Binder
cumulant~\cite{Binder81,Binder92}.  One can locate the phase
transition point by calculating, for example,
\begin{equation}
  U_4 = 1 - \frac{\langle \Phi^4 \rangle }{\langle \Phi^2
    \rangle\langle \Phi^2 \rangle},
\end{equation}
where $\langle \Phi^m \rangle$ denotes the $m$th moment of the
observable $\Phi$ (the order parameter) for different system sizes as
a function of the external parameter under scrutiny, and look for the
crossing point of these curves.  The essential point is that the
product of the powers in the numerator and denominator in the second
term are equal.  This method has even been applied to locate phase
transition points driven by quantum fluctuations~\cite{Hetenyi99}, but
only in cases where the order parameter is an expectation value of an
observable, rather than a Berry phase.  In this spirit, we calculated
$M_4/M_2^2$ (shown in Fig. \ref{fig:U4}).  On the attractive side, in
the insulating phase, this quantity has a negative slope as a function
of $V$ and exhibits size dependence, while in the conducting phase it
is constant and size independent.  On the repulsive no size dependence
is found until $V/t \approx 2$, but size dependence is found in the
insulating phase.  

\subsection{Total polarization distribution}

In Fig. \ref{fig:Px} we show the Fourier transform of $F_q$
($\tilde{F}_x$) as a function of the variable conjugate to $q$
(denoted by $x$) and the coupling $V$ for a system of size $L=30$.
Near $V=0$ $\tilde{F}_x$ is flat.  Structure begins to develop near
$V=\pm 2t$, but much slower on the positive side.  The structure
eventually consists of two sharp peaks in the region shown, one at
half of the lattice, the other at its edge.  The fact that there are
two is a clear consequence of the filling correction suggested by
Aligia and Ortiz~\cite{Aligia99}.  It is interesting that the behavior
of this quantity, which is related to the distribution of the center
of mass, behaves similar on both $\pm V$, even though the nature of
the ground states are very different.  In the extreme limits, $V
\rightarrow \pm \infty$ one can easily construct the probability
distributions for ground states (perfect CDW for $V \rightarrow
\infty$, all particles ``stuck'' together for $V \rightarrow
-\infty$).

To show this, let $N$ denote the number of particles, and $L$ the
number of lattice sites, and $L/N=2$.  At large and positive $V$ the
CDW state is doubly degenerate, we can write the expectation value of
$\hat{X}$ for each of these as:
 \begin{eqnarray}
  X_+ = \sum_{j=1}^N (2 j - 1) = N^2 \\
  X_- = \sum_{j=1}^N (2 j) = N^2 + N,
\end{eqnarray}
each occurring with a probability of one-half.  Since the operator
$\exp(i 2 \pi q \hat{X}/L)$ is diagonal in the position representation
we can write
\begin{equation}
  \label{eqn:Fq_cdw}
  F_q = \frac{1}{2}\left(e^{i\frac{2\pi q}{L}N(N+1)} + e^{i\frac{2\pi q}{L}N^2}\right) =
    \begin{cases}
      1 & \mbox{if $q$ is even} \\
      0 & \mbox{if $q$ is odd.}
    \end{cases}
\end{equation}

For the case of extreme clustering ($V \rightarrow -\infty$) the
ground state is $L$-fold degenerate.  The expectation values of the
total position are
\begin{equation}
  X_k = Nk + \sum_{j=1}^N j = N k + \frac{N(N+1)}{2},
\end{equation}
where $k$ corresponds to the $k$th degenerate ground state.  In this
case, $F_q$ becomes
\begin{equation}
  F_q = \frac{1}{L}\sum_{k=0}^{L-1} e^{i \frac{2\pi}{L} q \left[ N k + \frac{N(N+1)}{2}\right]}.
\end{equation}
It is easily shown, using that $N=L/2$ is odd that again we have
\begin{equation}
  F_q = 
    \begin{cases}
      1 & \mbox{if $q$ is even} \\
      0 & \mbox{if $q$ is odd.}
    \end{cases}
\end{equation}

\begin{figure}[ht]
  \centering
  \psfrag{x}{$x$}
  \psfrag{F}{$F$}
  \psfrag{V/t}{$V/t$}
   \includegraphics[width=\linewidth,keepaspectratio=true]{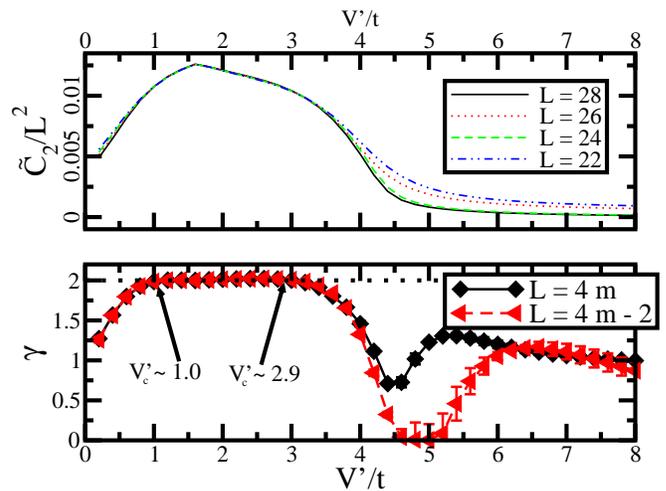}
   \caption{(Color online) Upper panel: $\tilde{C}_2$ as a function of
     $V'/t$ for a system of fermions with nearest and next-nearest
     neighbor interaction for the case $V=4t$.  Four system sizes are
     shown, $L=22,24,26,28$.  Lower panel: Size scaling exponent
     $\gamma$ for $\tilde{C}_2$ as a function of $V'/t$.  Diamonds
     show the results based on calculations for systems with sizes $L
     = 4 m$, where $m$ is an integer ($L=12,16,20,24,28$), left
     triangles show the results based on calculations for system sizes
     $L = 4 m - 2$, where $m$ is an integer ($L=10,14,18,22,26$).  The
     dotted line indicates the value of two.  $V_c' \approx 1.0$ is the
     critical point for the transition between the charge-density wave
     and the Luttinger liquid phases.  $V_c' \approx 2.9$ is the
     critical point for the transition between the Luttinger liquid
     and the bond-order phases. }
 \label{fig:gamV4}
\end{figure}

\section{Interacting model of spinless fermions with nearest and next nearest neighbor interaction}

The other model we study is an interacting model of spinless fermions
on a lattice in one dimension with Hamiltonian
\begin{equation}
  H = \sum_{i=1}^L [-t(c^\dagger_{i+1} c_i + c^\dagger_i c_{i+1}) + V n_i n_{i+1} + V' n_i n_{i+2} ].
\end{equation}
The phase diagram of this model was determined by Mishra {\it et
  al.}~\cite{Mishra11} In addition to the two phases already known (LL
and CDW) two more phases were found; a bond order (BO) phase at
intermediate $V'$ and another charge-density wave (CDW-2) phase for
large $V'$.  The latter consists of alternating pairs of particles and
pairs of holes.  As $V' \rightarrow \infty$ the ordered state that
emerges is one which pairs of particles alternate with pairs of holes.
We may write the four possible ordered states as
\begin{eqnarray}
  X_1 &=& \sum_{j=1}^{\frac{N}{2}} (4j-3) + \sum_{j=1}^{\frac{N}{2}} (4j-2), \\ \nonumber
  X_2 &=& \sum_{j=1}^{\frac{N}{2}} (4j-2) + \sum_{j=1}^{\frac{N}{2}} (4j-1), \\ \nonumber
  X_3 &=& \sum_{j=1}^{\frac{N}{2}} (4j-1) + \sum_{j=1}^{\frac{N}{2}} (4j), \\ \nonumber
  X_4 &=& \sum_{j=1}^{\frac{N}{2}} (4j) + \sum_{j=1}^{\frac{N}{2}} (4j-3).
\end{eqnarray}
Using these ordered states it can be shown that
\begin{equation}
  F_q = 
    \begin{cases}
      -1 & \mbox{if $q$ is even} \\
      0 & \mbox{if $q$ is odd.}
    \end{cases}
\end{equation}
We see that a sign change occurs between the two CDW phases (compare
with Eq. (\ref{eqn:Fq_cdw})) corresponding to a shift of the maximum
of the polarization.

In Fig. \ref{fig:gamV4} we present our results for $\tilde{C}_2$
(upper panel) and its finite size scaling exponent (lower panel) for
$V=4t$ as a function of $V'/t$.  In Ref. \onlinecite{Mishra11} it was found
that the CDW to LL transition occurs at $V_c' \approx 1.0t$, while
the LL to BO at $V_c'\approx 2.9t$ The BO phase transforms into the
second CDW phase at $V_c' \approx 4.29t$.  Other than the LL all other
phases are gapped and insulating.  In Ref. \onlinecite{Mishra11}
Fig. 3 shows the gap for this case, which is zero in the region $1.0t <
V' < 2.9t$, exactly where the scaling exponent $\gamma=2$
(Fig. \ref{fig:gamV4} lower panel).  Thus, phase transitions
accompanied by gap closure (metal-insulator transitions) can be
detected by our method.

While transitions between gapped to gapped phases are more difficult,
in this case the BO to CDW-2 transition can be approximately located
based on comparing calculations for systems of size $L = 4 m$ and $L =
4 m - 2$ ($m$ integer), since the ordered CDW-2 unit cell is four
lattice sites.  In the upper panel it is obvious that $\tilde{C}_2$
converges to different values for $L=24,28$ from those of $L=26,22$.
The two sets of curves start to deviate at $V' < 4.29t$, since it is
expected that the correlations associated with CDW-2 persist in the BO
phase.  It is interesting that the scaling exponents for $L = 4 m$ and
$L = 4 m - 2$ deviate significantly only after $V' \approx 4.3t$.

\section{Conclusion}

Even though the polarization in crystalline systems corresponds to a
Berry phase, proper finite size scaling is possible via discrete
formulas for gauge invariant cumulants.  The variance of the
polarization in hte Luttinger liquid (gapless) phase exhibit a finite
size-scaling exponent $\gamma=2$.  Based on this we were able to
identify metal-insulator transitions in several interacting models.
The main limitation in our study appears to be the small system sizes
accessible to exact diagonalization.

\section*{Acknowledgments}

This research was supported by the This research is supported by the
National Research, Development and Innovation Office - NKFIH within
the Quantum Technology National Excellence Program (Project No.
2017-1.2.1-NKP-2017-00001), K119442 and by UEFISCDI, project number
PN-III-P4-ID-PCE-2016-0032.


\begin{thebibliography}{9}

\bibitem{King-Smith93} R. D. King-Smith and D. Vanderbilt, {\it Phys. Rev. B}
  {\bf 47} 1651 (1993).

\bibitem{Resta94} R. Resta, {\it Rev. Mod. Phys.} {\bf 66} 899 (1994).

\bibitem{Berry84} M. V. Berry, {\it Proc. Roy. Soc. London} {\bf A392} 45
  (1984).

\bibitem{Xiao10} D. Xiao, M.-C. Chang, and Q. Niu, {\it Rev. Mod. Phys.} {\bf
  82} 1959 (2010).

\bibitem{Zak89} J. Zak, {\it Phys. Rev.} {\bf 62} 2747 (1989).

\bibitem{Bernevig13} B.~A. Bernevig and T.~L. Hughes, {\it Topological Insulators and Topological Superconductors} Princeton University Press, 2013.

\bibitem{Thouless82} D. J. Thouless, M. Kohmoto, M. P. Nightingale, and M. den Nijs
 {\it Phys. Rev. Lett} {\bf 49} 405 (1982).

\bibitem{Thouless83} D. J. Thouless {\it Phys. Rev. B} {\bf 27} 6083 (1983).

\bibitem{Fu06} L. Fu and C.~L. Kane, {\it Phys. Rev. B} {\bf 74} 195312 (2006).

\bibitem{Souza00} I. Souza, T. Wilkens, and R. M. Martin, {\it Phys. Rev. B} {\bf 62} 1666 (2000).

\bibitem{Resta98} R. Resta, {\it Phys. Rev. Lett.} {\bf 80} 1800 (1998).

\bibitem{Resta99} R. Resta and S. Sorella, {\it Phys. Rev. Lett.} {\bf 82}
  370 (1999).

\bibitem{Aligia99} A. A. Aligia and G. Ortiz, {\it Phys. Rev. Lett.} {\bf 82} 2560  (1999).

\bibitem{Nakamura02} M. Nakamura and J. Voit, {\it Phys. Rev. B}
  {\bf 65} 153110 (2002).

\bibitem{Oshikawa03} M. Oshikawa, {\it Phys. Rev. Lett.}
  {\bf 90} 236401 (2003).

\bibitem{Kohn64} W. Kohn, {\it Phys. Rev.} {\bf 133} A171 (1964).

\bibitem{Fisher72} M. E. Fisher and M. N. Barber, {\it Phys. Rev. Lett.} {\bf 28} 1516 (1972).

\bibitem{Binder81} K. Binder, {\it Phys. Rev. Lett.} {\bf 47} 693 (1981).
  
\bibitem{Binder92} K. Binder, {\it Annu. Rev. Phys. Chem.} {\bf 43} 33 (1992).
  
\bibitem{Kobayashi18} R. Kobayashi, Y. O. Nakagawa, Y. Fukusumi,
  M. Oshikawa, {\it Phys. Rev. B} {\bf 97} 165133 (2018).

\bibitem{Oshikawa18} H. Watanabe and M. Oshikawa, {\it Phys. Rev. X}
  {\bf 8} 021065 (2018).
  
\bibitem{Nakamura18} M. Nakamura, arXiv:1807.02864.

\bibitem{Lieb61} E. Lieb, T. Schultz, and D. Mattix, {\it
  Ann. Phys. (N.Y.)} {\bf 16} 407 (1961).

\bibitem{Hetenyi13} B. Het\'enyi, {\it Phys. Rev. B} {\bf 87} 235123
    (2013).

\bibitem{Patankar18} S. Patankar, L. Wu, M. Rai, J.~D. Tran, T. Morimoto, D.~E. Parker, A.~G. Grushin, N.~L. Nair, J.~G. Analytis, J.~E. Moore, J. Orenstein, D. H. Torchinsky, {\it P
  hys. Rev. B} {\bf
  98} 165113 (2018).

\bibitem{Yahyavi17} M. Yahyavi and B. Het\'enyi, {\it Phys. Rev. A} {\bf 95} 062104 (2017).
  
\bibitem{Chiappe18} G. Chiappe, E. Louis, J.~A. Verg\'es, {\it J. Phys. Cond. Mat.} {\bf 30} 175603 (2018).

\bibitem{Baeriswyl00} D. Baeriswyl, {\it Found. Phys.} {\bf 30} 2033 (2000).

\bibitem{Zak00} J. Zak, {\it Phys. Rev. Lett.} {\bf 85} 1138 (2000).

\bibitem{Marzari97} N. Marzari and D. Vanderbilt, {\it Phys. Rev. B}
  {\bf 56} 12847 (1997).

\bibitem{Marzari12} N. Marzari {\it et al.}, {\it Rev. Mod. Phys.} {\bf 84} 1419 (2017).

\bibitem{Hetenyi99} B. Het\'enyi, M. H. M\"user, and B. J. Berne, {\it Phys. Rev. Lett.} {\bf 83} 4606 (1999).

\bibitem{Mishra11} T. Mishra, J. Carrasquilla, M. Rigol, {\it Phys. Rev. B} {\bf 84} 115135 (2011).

\end{thebibliography}
\end{document}